\documentclass{www2010-submission}

\usepackage{subfigure}
\usepackage[usenames,dvipsnames]{xcolor}
\usepackage{verbatim}
\usepackage{algorithm}
\usepackage{graphicx}
\usepackage{multirow}
\usepackage{url}
\usepackage{alltt}
\usepackage{listings}

\begin{document}
%
\toappearbox{}
\title{ArcLink: Optimization Techniques to Build and Retrieve the Temporal Web Graph}
%
%
%
%
%

\numberofauthors{2} 
%
\author{
%
%
\alignauthor
       Ahmed AlSum\\
       \affaddr{Department of Computer Science}\\
      \affaddr{Old Dominion University}\\
       \affaddr{Norfolk VA, USA}\\
      \email{aalsum@cs.odu.edu}
\alignauthor
Michael L. Nelson\\
       \affaddr{Department of Computer Science}\\
       \affaddr{Old Dominion University}\\
       \affaddr{Norfolk VA, USA}\\
       \email{mln@cs.odu.edu}
   }


\maketitle

\newcommand{\todo}[1]{ {\textcolor{Red}{${\genfrac{}{}{0pt}{1}{TO}{DO}}$ {\bf #1}}} }

\begin{abstract}

 We present ArcLink, a proof-of-concept system that complements open source Wayback Machine installations 
 by optimizing the construction, storage, and access to the temporal web graph. We divide the web graph construction into four stages (filtering, extraction, storage, and access) and explore optimization for each stage. ArcLink extends the current web archive interfaces to return content and structural metadata for each URI. 

\end{abstract}



\section{Introduction}

The web graph is a directed graph where each vertex represents a URI and the edges represent a hyperlink between them. The web graph was the foundation of various applications; ranking such as: PageRank \cite{Brin1998} and Kleinberg HITS \cite{Kleinberg1999}, web spam detection \cite{Fetterly2004a,Becchetti2008},  finding  related pages \cite{Dean1999}, and building web graphs based on top-level domains and not individual pages \cite{Bharat1998}.

\begin{figure*}[htb]
	\centering
	\subfigure[Regular WG.]{
		\includegraphics[width=1.2in]{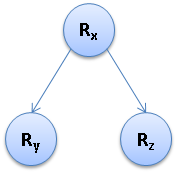}
	\label{fig:wg}
	}
	\subfigure[WG at t1.]{
		\includegraphics[width=1.2in]{WG.png}
		\label{fig:twgt1}
	}
	\subfigure[ WG at t2.]{
		\includegraphics[width=1.2in]{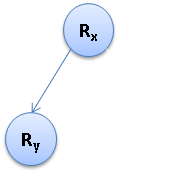}
		\label{fig:twgt2}
	}
	\subfigure[Temporal WG.]{
		\includegraphics[width=1.3in]{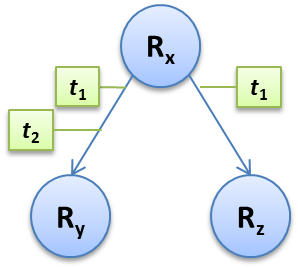}
		\label{fig:twg}
	}
\caption{Regular and Temporal Web Graph.}
\label{fig:WGTWG}
\end{figure*}

Web archives preserve web pages before they change or disappear forever. Each page, identified by a URI, may have one or more snapshots (called mementos) with different timestamps from which we can construct a Temporal Web Graph.
The temporal web graph may have more than one memento for each URI. 
Figure \ref{fig:wg} illustrates the regular Web graph. Each URI is connected with another URI by only one link because there is only one snapshot for the source URI. In the temporal Web graph (figure \ref{fig:twg}),
some of the outlinks changed through the time. At $t_1$, $R_{x}$ has two outlinks to $R_{y}$ and $R_{z}$ (figure \ref{fig:twgt1}), then at $t_{2}$, $R_{x}$ has only one outlink to $R_{y}$ (figure \ref{fig:twgt2}).
  So the graph structure is changing through the time.

The paper uses Memento protocol notation 
 \cite{VandeSompel2011}. Memento is an extension for HTTP protocol to allow the user to browse the past
web as the current web. \textit{URI}-\textit{R} denotes an original resource that exists or used to exist in the live web, and \textit{URI}-\textit{M} identifies the memento that is a snapshot for the original resource as it appeared in the past and preserved by the web archive. 
The time that the memento was observed (or captured) by the archive is known as \textit{Memento-Datetime}.

\subsection{Motivation}

The International Internet Preservation Consortium (IIPC)\footnote{\url{http://netpreserve.org}} was formed in 2003 to improve the tools, standards, and best practices in web archiving. 
The IIPC Access working group proposed a set of use cases for accessing web archives \cite{iipc2006}; these use cases covered various types of users. The report highlighted the importance of the linking information (the outgoing and the incoming links) of the archived web  for analysis purposes. The report suggests archives provide an API interface to query the link structure for specific URI, while recognizing that providing such an API is time-consuming. 


The time dimension in the web graph extends the traditional web graph application. For example, the temporal web graph could be used in ranking the full-text search for the web archives related to the time. 
Also, the results should differ based on the time filter\footnote{\url{http://www.alexa.com/help/traffic-learn-more}} (e.g., the query string ``Email'' on 2004 might bring ``Yahoo Mail'' as the first result, today, the same term may return ``Gmail''). 
The temporal web graph could be used by the web crawler to discover new URIs, the increasing number of inlinks could prioritize the crawling schedule. Researchers can use the temporal web graph to study the evolution of the web.

The web archive does not cover everything in the past, the archival coverage differs based on the popularity of the URI \cite{Ainsworth2011}. So the Temporal Web Graph will suffer from the missing nodes problem. 
For example, if $URI-R_{x}$ has $n$ webpages that pointed to it ($n$ inlinks), the web archive may miss part of $n$ inlinks. 
The number of incoming links for  $URI-R_{x}$ is determined by the available mementos that point to this URI which may be different from the live web inlinks. So even though $URI-R_{x}$ may be a popular URI and was captured successfully, the web archive may not archive all the related URI-R that pointed to it and may underestimate the importance of $URI-R_{x}$. 

The Internet Archive (IA)\footnote{\url{http://www.archive.org}} is the largest and the oldest web archive in the world. AlNoamany et al. \cite{Alnoamany} found that more than 90\% of the sessions and over 50\% of the requests to the IA's Wayback Machine \cite{tofel2007} came from robots.
This high percentage of robots in the archives suggests that there is significant demand for content and link structure metdata about mementos and a significant portion of the current load on the IA's Wayback Machine could be offloaded to an API.


\subsection{ Current Approaches}
Interaction with the past web  is limited to displaying the webpage as it appeared in the past. The interface may support URI lookup (e.g., Internet Archive WayBack Machine) or full-text search typically on focused collection (e.g., the British Library and Archive-It). The web archives interfaces do not provide an access to the archived web metadata such as: titles, thumbnails, and link structure. 
Web archives do not currently support the rich ecosystem of APIs, such as twitter or facebook.
 CommonCrawl\footnote{\url{http://commoncrawl.org}} provide an access to a set of metadata (including outlinks) using Amazon EC2 in flat file format.

Public users and researchers could only scrape the archived pages and do the extraction for the links and other metadata. Br\"{u}gger \cite{Brugger2012} used IssueCrawler\footnote{\url{https://www.issuecrawler.net/}} to crawl the Internet Archive and national Danish web archive Netarkivet\footnote{\url{http://netarkivet.dk/}} to analyze the hyperlink network in web archives; Br\"{u}gger showed the inadequacy of the crawler to work on the web archive.
Weber \cite{Weber2012} built a custom web crawler, called HistoryCrawl, to crawl the Internet Archive interface to extract the hyperlinks through the time.
Klein and Nelson \cite{Klein2010} had to download and page scrape every memento in a TimeMap and extract the changing titles in html pages. Also, Padia et al. \cite{Padia2012} downloaded pages from Archive-It collections to extract terms for new visualization techniques.
In this paper, we show the bad performance of this technique on both the web archive resources and the extraction performance.

%
%
%

\begin{figure*}[btp]
\centering
\includegraphics[width=5in]{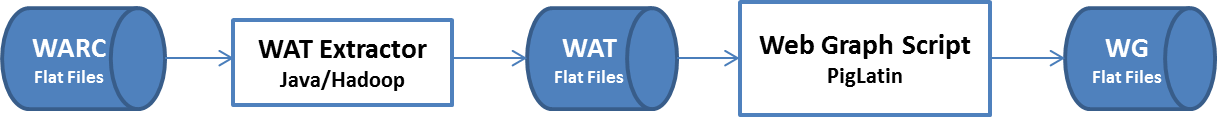}
\caption{Internet Archive Web Graph extraction architecture.}
\label{fig:iadiagram}
\end{figure*}

The Internet Archive has built a web graph for some collections for internal use where  regular users do not have access to it. IA has two-phase extractor to construct the web graph. Figure \ref{fig:iadiagram} illustrates the IA web graph extraction architecture.
First, IA has developed a new metadata file called \textit{Web Archive Metadata file} (\textit{WAT})\footnote{\url{https://webarchive.jira.com/wiki/display/Iresearch/Web+Archive+Metadata+File+Specification}}. WAT is a metadata file format that carries metadata about the WARC file  \cite{warciso} including the outlinks. They extracted the metadata directly from the WARC files using Java programs.
The second phase used the WAT files to construct the web graph. IA used PigLatin scripts to extract the complete web graph from the outlink field in the WAT file.
 The limitations of this technique are:
 \begin{enumerate}
\item  Generally, it is a batch process, and can not be implemented on a finer-grained level (e.g., site or domain). 
\item  IA used the final web graph internally. There is no public access point to reach this data. 
\item  The extracted web graph can not be aggregated with other web graphs on the same archive or other archive. 
\item  The  technique does not support incremental update.
\end{enumerate}

In this paper, we propose new optimization techniques for the creation, preservation, and retrieval for the web graph considering the time dimension. The contribution started with decreasing the size of the data, hashing technique for the URI to enable a native distributed processing with linear and equal insertion and update overhead, efficient schema to represent the time dimension, and API interface for accessing the web graph. ArcLink is a complete system using the Hadoop\footnote{\url{http://hadoop.apache.org/}}  framework that implements these optimization techniques. 

\subsection{DataSet}
 We used the collaborative IIPC Olympics 2010 collection\footnote{\url{http://olympics.us.archive.org/}},  a collection of websites about the 2010 Winter Olympics for experimental use. Twelve institutions contributed seeds for the project. Internet Archive did the crawls and made accessible all the harvested materials including extracted metadata and crawl reports. 
The collection was crawled between 11/2009 and 03/2010 during the 2010 winter Olympics. 
The corpus size is 700+GB, the crawler started with a seed list of 302 websites. 
 Table \ref{tab:olympicsstat} lists more information about the collection.
\begin{table}
\center
\caption{IIPC 2010 Olympics Collection.}

\begin{tabular}[h]{l|r}\hline
Size & 700+GB \\ \hline
Number of WARC files & 1950 \\ \hline
Number of Seed URI-R & 302 \\ \hline
Number of Unique URI-R & 6.4 M\\ \hline
Number of URI-M & 23.7 M\\ \hline
Start date & Nov 3, 2009 \\ \hline
End date & Mar 13, 2010 \\ \hline
\end{tabular}
	\label{tab:olympicsstat}
\end{table}

\section{Related Work}

Link Database \cite{Randall2002}, a part of the Connectivity Server \cite{Bharat1998a} (used by AltaVista),  provides a fast access storage to the web graph. Link Database compressed the link id to 6 bits per link, so the graph could be loaded into the main memory. 
Scalable Hyperlink Store (SHS; used by Microsoft) \cite{Najork2009} is a distributed in-memory ``database'' to store the web graph. SHS provided fast access by keeping the web graph information in the main memory. SHS provides a kind of API to facilitate the interactions with SHS servers. 
Suel and Yuan \cite{Suel2001} created a hostname list and URI list. They used Huffman codes to compress each one, then divided the links into global links between pages on different hosts, and local links between pages on the same host. These systems were built to run on a single machine, and mainly to work on the live web; they did not take the temporal dimension of the web archive in consideration.

Avcular and Suel \cite{Avcular2011} discussed distributed manipulation of archival web graph using Hadoop. 
MapReduce \cite{Dean2004} is a distributed programming framework to process large scale data. Hadoop 
is the open source implementation for MapReduce.
Pregel \cite{Malewicz2010} is a distributed system for efficient processing the large scale graph. 
PeGaSus \cite{Kang2010} is a petascale graph mining library to process the large web graph. 
 Donato et al. \cite{Donato2004} studied the properties of web graph based on Stanford WebBase collection \cite{Cho2006}. 

 Bordino et al. \cite{Bordino2008} provided statistical analysis to the temporal characteristics of 100M mementos 12 months snapshots of the .uk domain  captured between June 2006 to May 2007. These built a web graph for each month, the web graph has been built using Web Graph \cite{Boldi2004}.
 
Anand et al. \cite{anand2009everlast} proposed EverLast, a distributed framework to address the challenge of capturing, preserving and querying the web archive. EverLast proposed Time-Travel Index (TTI) to support the query on the web archive.
Song and JaJa \cite{song2008fast} proposed EdgeRank technique to partition the web Graph in order to merge the web pages in container. The new algorithm decreased the number of containers that are needed to browse the web archive that enable faster browsing of archived web contents. Also, Song and JaJa \cite{Song2008} proposed a new schema for efficient preservation and retrieval for the archived web.


%
%
%
%
%

\section{ArcLink Stages}
ArcLink  is a distributed processing system to pre-process the link structure information from the archived web collections and create a temporal web graph. ArcLink has four main stages: 
filtering, extraction, storage, and access. 

\begin{figure*}[t]
	\centering
		\includegraphics[width=5.5in]{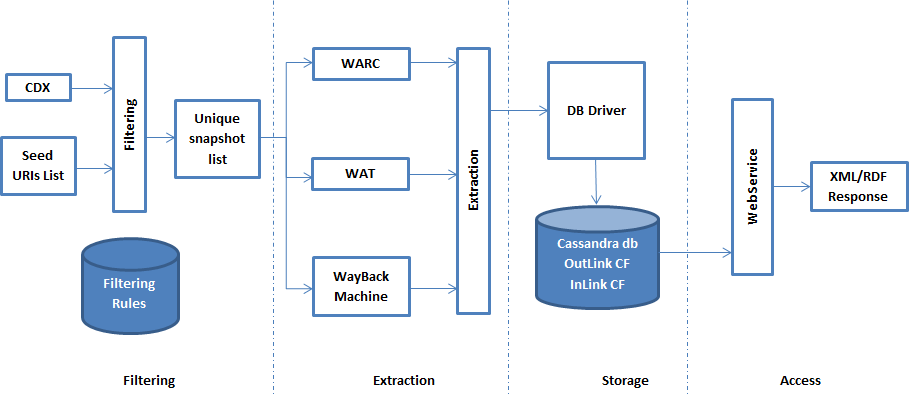}
	\caption{ArcLink Architecture}
	\label{fig:arclink_diagram}
\end{figure*}

Figure \ref{fig:arclink_diagram} illustrates the different stages and the relation between them.
\textit{Filtering} is using the crawler log to omit the mementos that will not contribute to the final temporal web graph. The goal is to reduce the size of the input corpus to optimize the following stages. \textit{Extraction} is responsible for extracting the outgoing links from the mementos. Extraction will compare between various input sources. \textit{Storage} is preserving the web graph information into a database for further usage. \textit{Access} will provide APIs interface that has the outgoing and incoming links for the requested URI.

In this paper, we studied the characteristics of each stage, and proposed the suitable optimization approaches. 

\subsection{Filtering Optimization}

The goal of the filtering optimization is to reduce the size of the input corpus to focus on the snapshots that carry link structure information. Lee et al.  \cite{Lee2007} defined a page that has no outlinks as a dangling page and proposed how it should be incorporated in the PageRank computation. ArcLink avoides the dangling nodes in the filtering phase, and may be included later if ArcLink detected a link to this dangling node.
The input for the ArcLink system is a list of the URIs within a collection. For Heritirix\footnote{\url{http://crawler.archive.org/index.html}}-based web crawlers  
, the URI list could be found in the crawler log entitled ``CDX file''\footnote{\url{http://archive.org/web/researcher/cdx_file_format.php}}. The CDX file is a space delimited file, each record belonging to one memento. The information includes the URI-prefix, timestamp, URI, mimetype, response code, page checksum, redirection URI, offset, and WARC file name. 
 
\footnotesize
\begin{verbatim}
example.org/foo.html 20090312223142 
   http://www.example.org/foo.html text/html 200
   E6T72C2R6BRRKSI3IZPMRJDXTFJIRC7P - - 111739 
   TENN-000001.warc.gz
\end{verbatim}
\normalsize
ArcLink filtering optimization techniques use this information to create a unique list of mementos that will contribute to the web graph. Reducing the input size will improve the extraction time and the required storage space.



\subsubsection{Filtering rules}\label{sec:filteringrules}
Building the web graph started with extracting the outlinks of the web page and creating the inlinks model from that. Based on this procedure, the filtering stage will exclude any snapshot that does not have outlinks (e.g., images), the following set of filtering rules could be used:
\begin{itemize}
\item HTTP Status: Include memento with a successful HTTP status (e.g., 200).
\item Content Mimetype: Exclude mementos with mimetype that do not carry textual content (e.g., images, javascript, stylesheets).
\item Resource extension: Exclude mementos that end with non-textual contents even if the mimetype is text/html. 
\item Content checksum: CDX file has the page content checksum calculated with SHA-1 algorithm for each memento. This value could be used to exclude the duplicate mementos.
\end{itemize}

\subsubsection{Implementation}
Apache Pig 
is an Apache open source tool that is customized to parallel processing of a large scale data. 
Filtering rules are written with PigLatin script that is capable of loading the CDX file fields, then applied INCLUDE/EXCLUDE filters. 
The output of the filtering steps is a list of the unique mementos after applying the filtering rules.

\subsubsection{Experiment }
To quantify the efficiency of the filtering optimization techniques, we ran different filtering rules to the Olympics 2010 collection CDX files. The input file has 23.7M URI-Ms. The success criteria was how to include the mementos that may contribute to the temporal web graph and how to exclude the dangling and the duplicate nodes that may not add any value to the web graph. The experiment will calculate the reduction in the number of items by applying each rule and what is the total gain of applying all the filters. The efficiency will be the percentage of mementos in the output snapshot list to the original number of snapshot. 
\begin{table*}[tb]
	\caption{Reduction Efficiency Experiment Results.}
	\centering
		\begin{tabular} {l  p{5cm} r r r r} \hline
			\textbf{Rule type} & \textbf{Rule parameter}	&	\textbf{Reduction in \#Records (23.8M)}	&\textbf{Time}	\\ \hline
	INCLUDE &	HTTP Status 200		& 17.908M (75\%) 	&936 sec			\\
	EXCLUDE &	Images, JS, and CSS & 	15.922M (67\%) &	807 sec	\\
	INCLUDE	&	text/* only					& 	12.656M	(53\%)&  744 sec		\\
  EXCLUDE	&	Resources with image extension		&	16.338M (69\%)	& 845 sec				\\
	EXCLUDE	&	Duplicate checksum			&	9.191M (39\%)	& 886 sec				\\  \hline
			\multicolumn{2}{c}{All the rules}		&	6.789M (29\%)	& 2098	sec		\\ \hline
		\end{tabular}
	\label{tab:filteringExp}
\end{table*}

\subsubsection{ Results and Analysis}
The filtering stage used pre-known information that was addressed by the crawler during the capture time to save the computation time that may result in unuseful (e.g., URIs do not have links) or duplicate content. 

Table \ref{tab:filteringExp} shows the experiment results. First, only 75\% have HTTP status 200; this means 25\% of the collection did not have an available representation or carried HTTP redirection status. 
Second, 33\% of the collection was for embedded resources (e.g., images, stylesheets). Third, the percentage of duplication in the collection is approximately 61\%. Finally, applying all rules showed that the efficiency was so significant to reduce the input records to 29\% of the original size. This reduction of the number of records reflected directly into the computation time.

The filtering rules approach is flexible enough to manage several kinds of computation. For example, it could be used to extract images only or videos only by adapting rules to filter by mimetype.

Generally, the crawling log is not available for the public users.
The screen scraping technique could not benefit from this information which added extra load to the web archive and the client. We recommend the web archives to publish this information to the public to help the third-party developer/researcher. IIPC funded the IIPC Memento Aggregator Experiment\footnote{\url{http://netpreserve.org/sites/default/files/resources/Sanderson.pdf}} to aggregate all the CDX files of the distributed archives of IIPC members to provide Memento based access to the holding of the open/restricted/closed archive.

\subsection{Extraction Optimization}
The first step of constructing the web graph is extracting the outgoing links from the web pages.
Optimizing the extraction stage focused on two things: the creation of the URI-ID, and extraction mechanism (data source and tools).

\subsubsection{URI-ID Generation}
The creation of a unique ID for each URI is a common technique in the web graph creation. 
The ID creation approaches may depend on ordering, by lexicographical ordering \cite{Randall2002,Najork2009,Avcular2011,Boldi2004,springerlink:10.1007/3-540-45703-8_30}, inlinks degree \cite{Adle} then applying Huffman codes, or storing in array then using the index as ID \cite{Bharat1998a}. These approaches made a tightly coupled relation between the URIs themselves, so the parallel and distributed processing were impossible. Also, it affected the update/re-indexing process.  
To avoid these problems, ArcLink generates a unique ID for each URI as the following:
\begin{enumerate}
\item Canonicalized the URI into SURT  format\footnote{\url{http://crawler.archive.org/articles/user_manual/glossary.html#surt}}.

\begin{small}
$\left. 
  \begin{array}{l l}
   www.example.org/foo.html \\
   example.org/foo.html\\
	www1.example.org/foo.html\\
  \end{array} \right\}\rightarrow org,example)/foo.html$
\end{small}

\item ArcLink encodes this canonicalized SURT string using SimHash \cite{Charikar2002} with a length of 128 bits. 
\end{enumerate}
Using this 1-1 mapping between the URI and the ID gives the ability to incremental and distributed processing for the same URI on different cycles or different machines because the ID generation depends on the URI only. 

\subsubsection{Data sources}
The extraction of the link structure from the archived web data differs from the extraction from the live web in the nature of the format of the archived web. ArcLink can extract the link structure from three sources based on the availability of the input:
\begin{enumerate}
\item \textit{Web ARChive file} (\textit{WARC}): 
 is the standard file format for web archives that offers a convention for concatenating multiple resource records (data objects), each consisting of a set of simple text headers and an arbitrary data block into one long file. 
\item \textit{Web Archive Metadata file} (\textit{WAT})
: is a new metadata file format that carries metadata about the WARC file including the outlinks. 
\item \textit{Web Archive UI}: that displays the web page as it appeared in the past. It is mainly used through the screen scraping techniques.
\end{enumerate}

\subsubsection{Implementation}
The Apache Hadoop is a framework that allows for the distributed processing of large data sets across clusters of computers using a simple programming model. We built MapReduce jobs using Java, the mapper part started with one memento at a time and extracted the outlinks based on the available source. For extracting the link structure from the WARC and web interface, we used HTML Parser\footnote{\url{http://htmlparser.sourceforge.net/}}, is a Java library used to parse HTML.
 ArcLink focused on the following items: the outlink (anchor link or embedded resources link), the type (href or image), and the associated text (anchor text for the href or the alternate text for the image). The reducer is responsible for ID creation for each of the extracted links and creating one record that contains (Document Checksum, Outlink URI, OutLink ID, type, text).

\subsubsection{Partitioning}
 Hadoop provides a transparent partitioning technique to divide the data on the cluster machines, but 
 the input file for the extraction does not carry the input data, it only had pointers to the actual data (e.g., WARC files).
Each line in the input file has the URI and WARC file name with an offset to the content. Processing each record requires an  access to the file on the harddisk. The simultaneous multi-access to the same file affects the performance of the extraction stage \cite{Najork2009}. ArcLink uses unix commands (sort and split) to partition the input files to ensure single access to the WARC file per task. Each WARC file name should appear in only one split file, the input split file could contain one or more WARC file. Each split file will be assigned to one Map task.

\begin{table*}[tbh]
	\caption{Extraction Experiment Results.}
	\centering
		\begin{tabular} {l| l| r  rr} \hline
 \textbf{Input}& &  \textbf{ Map   } & \textbf{  Reduce   }& \textbf{ Total time }\\ \hline
\multirow{2}{*}{2 Tasks (Partition)}&\textit{WARC} &  13,327 & 2,770 &16,098 \\ 
									 &\textit{WayBack Machine} & 21,422 &4,194  & 25,616\\  \hline
 
\multirow{2}{*}{2 Tasks (Normal)}& \textit{WARC} &  15,324 & 2,940 & 18,265 \\  
								   & \textit{WayBack Machine} & 17,447 &  4997& 22,444\\  \hline
\multirow{2}{*}{5 Tasks (Partition)}&\textit{WARC} & 8,304 &  1,746 & 10,051 \\ 
									 &\textit{WayBack Machine} &13,721 &2,257  & 15,978\\  \hline
		\end{tabular}
	\label{tab:extractionExp}
\end{table*}

\subsubsection{Extraction from WAT files}
The evaluation does not cover the extraction from the WAT files. Theoretically, the extraction from the WAT file is faster than the extraction from the actual content (WARC files) because the WAT file has been pre-processed earlier. The WAT creation is an expensive task because it extracts a lot of information from the actual text. So the quantitative comparison should take into the consideration the time  to create the WAT file. Also,  WAT files are not ready for all the collections. Finally, WAT file does not have all the required information that were addressed in this system.


\subsubsection{Experiment }
In this section, we perform a quantitative study to compare the various extraction techniques.We extracted the outlinks from two sources (WARC, Wayback Machine) for the same collection. We repeated the experiment on different number of MapReduce jobs. We have two samples of data. For the two tasks experiments, we sampled 800k records from the filtering phase output, and we feed the extractor on two modes: \textit{Normal} using the default Hadoop partition, and \textit{Partition} in which ArcLink splits the data before the submission. For the five tasks run, we sampled 500k records. The task has been repeated to extract from the WARC file and WayBack Machine. Accessing  the WayBack Machine was done without any politeness period between the requests.
%

 Table \ref{tab:extractionExp} shows the results for \textit{Map phase} (total time for all mappers), \textit{Reduce phase}, and the total time for both in seconds. The results showed that using WARC as data source is 61\% efficient than performing page scraping from the WayBack machine interface. Also, the partitioning technique has  better performance with two tasks (85\%), increasing the number of tasks made the partition enhancement significant.

Extracting from the Olympics 2010 dedicated Wayback Machine server failed in the first round because of the server was not prepared to receive the resulting high load. After updating the memory configuration, we were able to access the server with different tasks. This problem does not happen with WARC extraction because we depend on Hadoop DFS (part of Apache Hadoop project) which was designed to receive high load of requests. It showed the negative impact of the page scraping on the archive resources. 

\subsection{Storage Optimization}
In the previous stages, the optimization techniques focused on the time and the computation power. In the storage optimization, ArcLink will optimize the required space to store the web graph. ArcLink preserves the extracted links structure for future access using a database which will be the repository of the link structure information for any further experiments. 

\subsubsection{Schema}
The outlinks and inlinks with temporal dimension is many-to-many relation with various properties. For example, the URI-R may have different mementos, each memento may have the same or different outlinks with different anchor text.

The main research question is how to represent the web graph with the temporal information. Usually, the web graph is represented as a directed graph: the vertex represents the URI and the edges represents hyperlinks between the URIs. With the temporal dimension, $URI-R_{x}$ could point to $URI-R_{y}$ at different timestamps with different properties (i.e., anchor text). Adding this information to the graph needs more sophisticated method to make the graph with the minimum size. We created  two schemas to represent the temporal web graph. 

\begin{figure*}[tb]
\begin{center}
\subfigure[web graph with temporal properties.]{
\includegraphics[width=3.2in]{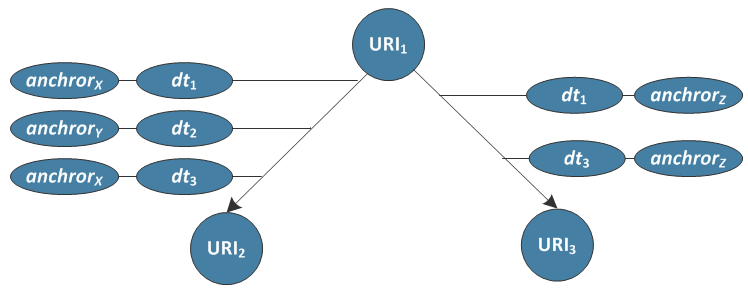}
\label{fig:tempgraph-properties}
}
\subfigure[Content-centric.]{
\includegraphics[width=1.3in]{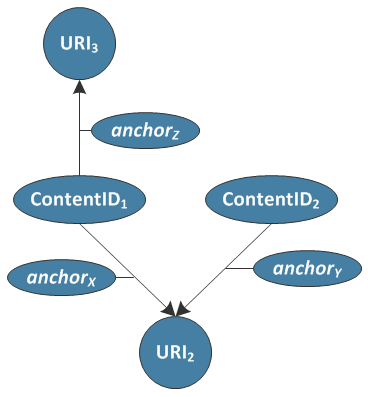}
\label{fig:tempgraph-content}
}
\caption{Temporal Web Graph approaches.}
\label{fig:tempgraph}
\end{center}
\end{figure*}

\begin{itemize}
\item \textit{Web Graph with temporal properties} (figure \ref{fig:tempgraph-properties}): In this schema, we expanded the regular web graph to include attributes for the edges. For each edge (that represents link from $URI-R_{x}$ to $URI-R_{y}$), we added two fields, the datetime for this memento and the anchor text for this references. We used this schema for access (section \ref{ch:access}), because it is more readable for the users and the applications.

\item \textit{Content-centric temporal web graph} (figure \ref{fig:tempgraph-content}): In this schema, we replaced the $URI-R$ and datetime attribute with the checksum for the content for this memento. The duplicate mementos have the same content (with same checksum) which evolve to the same vertex. This schema is used for the preservation, because focusing on the content will remove the duplicate information. The rest of this section will explain the schema implementation in Cassandra db.
\end{itemize}

\subsubsection{Implementation}
ArcLink reference implementation is using the Apache  Cassandra\footnote{\url{http://cassandra.apache.org/}}
database which is a highly scalable, distributed, and structured key-value store. We used the Super Column Family structure to build the schema for saving the link structure information. The advantage of this technique is providing the same analogy of the temporal relation between the datetime and the list of mementos with different attributes for each one. The Cassandra db handles the update/insert operations, so even if we insert the same record twice it will detect it was previously inserted and update the content with the new information.


\subsubsection{Experiment and Results}\label{ch:storageExp}
In this section, we evaluate the efficiency of the new schema in both space (reduction in storage) and time (insertion/update).

\begin{figure*}[tb]
\begin{center}
\subfigure[Insert Time.]{
\includegraphics[width=3.0in]{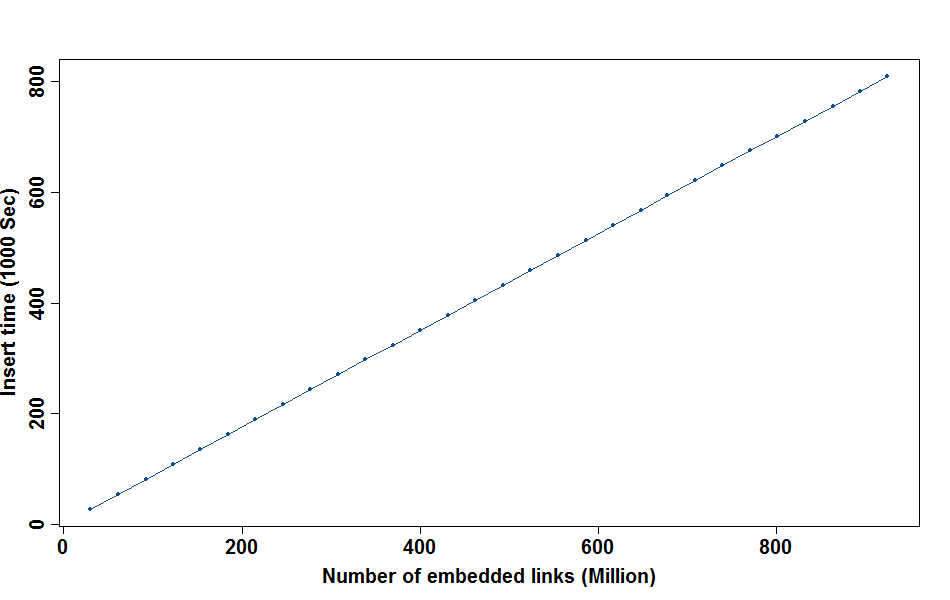}
\label{fig:dbInsert}
}
\subfigure[Update Time.]{
\includegraphics[width=3.0in]{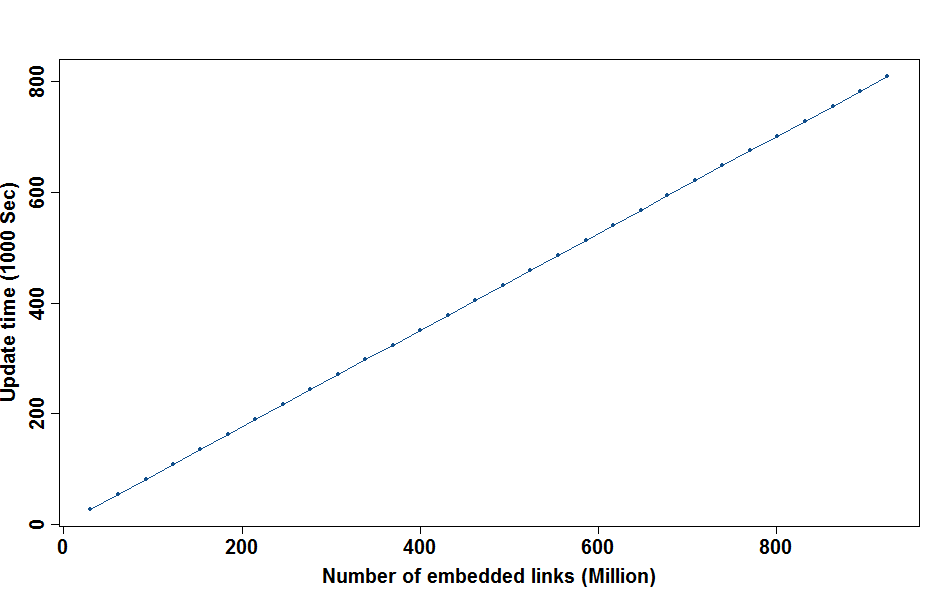}
\label{fig:dbUpdate}
}
\caption{Insert/Update time.}
\label{fig:storageEvaluation}
\end{center}
\end{figure*}

The database driver is the program that is responsible of inserting the extracted links into the database, creating outlinks and inlinks tables. The Cassandra database driver calculated the required time for the insertion. Then, we repeated the process again to calculate the time for the update. Figure \ref{fig:storageEvaluation} shows that there is a linear relationship between the number of links and required time for insertion. Also, the same linear relationship appeared for the update which means there is no extra overhead for the update process.

The content-centric schema focused on the content more than the URI. The same checksum may belong to one URI in one timestamp (unique memento), one URI in different timestamps (duplicate mementos), or different URIs in different timestamps (duplicate content).  We found duplicate content was common, for example, the Olympics collection has 23K+ mementos that have custom ``Page not found'' responses that return 200 instead of 404  \cite{Bar-Yossef2004}. In the regular web graph, each one of these snapshot will have one vertex that has the same outlinks. 

\subsection{Access}\label{ch:access}
ArcLink provides APIs to enable the  applications to access the link structure information. ArcLink delivers the link structure information to other systems instead of processing/analyzing the information by itself. It makes  ArcLink  a source of knowledge which could be expanded by incrementally processing more archived web material or by aggregating the ArcLink interface with other archives. ArcLink delivery method depends on REST web services. ArcLink response is annotating the URI-R with the outlinks and inlinks predicates through the time. Listing \ref{lst:accessschema} shows a complete request/response session for ArcLink. Line 1 shows the curl request, the service take a \texttt{uri} parameter that denotes the requested URI. Line 3-32 show the response in RDF/XML. Line 5 shows URI as  subject. Lines 6-18 show the \textit{hasOutlinks} to list the outlink set with the observed datetime and anchor text. Lines 19-31 shows \textit{hasInlinks} to list the inlink set.

ArcLink interface is built on RDF that could be aggregated with other ArcLink instances that  carry information about the requested URI-R. The aggregation could be done on the response level to give each ArcLink implementer the freedom to adjust the implementation details based on the requirements and capabilities. 

\section{Scalability and Cost Model}
ArcLink has been designed to deal with large-scale web archives. The evaluation for the different stages showed linear relation in every stages. Filtering stage applied different rules to the CDX input file; each record was visited once. The total complexity of this stage is O(n) where n is the number of records. Extraction  was also  linear; each file is visited only once. Section \ref{ch:storageExp} discussed the linear complexity for the storage stage.
Additional to the construction process, the access stage could be aggregated with other web graphs based on the RDF aggregation schema.
Based on our empirical study, we present the following cost model.
\begin{itemize}
\item 
\textbf{Filtering stage}\\
Assume you have CDX input file with size $n\;mementos$ 
using $m$ machines in Hadoop cluster.
\begin{equation}
FilteringTime = n/10^6 * 88/ m\;(sec)
\end{equation}
\begin{equation}
FilteringReduction=n * 0.30\;(mementos)
\end{equation}

\item
 \textbf{Extraction stage}\\
Assume you have a web archive corpus with $n\; mementos$ using $m$ machines in a Hadoop cluster.
\begin{equation}
ExtractionTime = n/10^6 * 5.5 / m\;(hrs)
\end{equation}

So you can extract the link structure for 1 billion mementos with in 2 days using a cluster of 100 nodes.

\item 
\textbf{Storage stage}\\
In the previous stages, all the output data is temporary and consumed by the next phases. The storage stage will preserve the link structure for further access. The space here is a major issue.
The required space for the Cassandra database was 35GB for outlinks and almost the same for inlinks. Also, it consumed 2GB for the total URI ID information. So the total required space was almost 10\% of the original size of the collection. Extraction stage produced the web graph in a flat file with size 120GB.
\begin{equation}
StorageSize=Out(n)+In(n)+Link(n)
\end{equation}
$Out(n)=n*0.05;In(n)=n*0.05;Link(n)=n*0.002$
$where\; n \;$ is the size of the collection.

\end{itemize}
If we were to scale this up to the recently announce WayBack Machine with 240B mementos and the size reached 5PB\footnote{\url{http://blog.archive.org/2013/01/09/updated-wayback}}, the results would be as the following:

 $FilteringTime= 240*10^9/10^6*88 /100= 58.6\;hrs$

 $FilteringReduction=240*10^9 * 0.30 = 72 *10^9 \;mementos$

 $ExtractionTime= 72,000 *10^9/10^6 * 5.5  / 100 = 165\;days $

 $StorageSize = 5\;PB * 0.05+5*0.05+5*0.0.2 = 500\;TB$


\section{Applications}
\subsection{Temporal Web Graph Properties}

The temporal web graph for Olympics 2010 collection contributed 19.8M nodes (each node represented URI-R) with 792+M edges (each edge represents a hyperlink). We analyzed the URI-R set and found that 18.57M did not have any mementos in the collection. So for this bounded collection, 92\% of the outlinks are not crawled which means less ability to the user to browse the collection as it appeared in the past.

\subsection{Temporal PageRank}
PageRank \cite{Brin1998} is a ranking technique that is used to rank a set of web pages based on the link structure. In this section, we tested the ranking information through the time. First, we calculated the page rank for the whole collection without the time dimension. We create an edge between $URI-R_{x}$ and $URI-R_{y}$ if $URI-R_{x}$ pointed to $URI-R_{y}$  in any point of the time. Then, we repeated the algorithm with the time dimension. We divided the graph per month. 
For example, for January 2010 graph,  we created an edge between $URI-R_{x}$ and $URI-R_{y}$ if $URI-M(URI-R_{x})$ during Jan 2010 pointed to $URI-R_{y}$.

Table \ref{tab:rank} shows the PageRank for the  collection per month. The last column has the rank for the whole collection. The rank for each month is affected by the crawling activity during this month. November 2009 focused on the preparation of the Olympics, the \url{vancouver2010.com} domain has the highest rank during this month. By March 2010, the news about the Olympics became more important. Newspaper sites like \url{lefigaro.fr} and \url{lemonde.fr} got higher ranks. We could consider the whole collection rank as the intersection of the rank for the various months. 
We calculated the overlapping (the intersection between two sets) and correlation (using Kendall $\tau$)  for the top 50 results in each month as described in \cite{McCown2007}. We compared between each subsequent months, then with the whole collection results. Table \ref{tab:cor} showed a weak relation between the ranks in different months, that means we can not depend on specific time rank to estimate another time period.

The temporal PageRank is affected with the available mementos on this date. We were not able to get a page rank for December 2009 because there were so few mementos captured on that month.

\begin{table*}[tbh]
	\caption{Collection Ranking Through the Time.}
	\label{tab:rank}

	\centering
\begin{tabular}{r| p{2.3in}|p{1.6in}|p{2.3in}}\hline\hline
&\textbf{Nov-2009} & \textbf{Dec-2009} & \textbf{Jan-2010} \\ \hline\hline
1&vancouver2010.com/code & - & topsport.com/sportch/liveticker/ \\
2&vancouver2010.com/en/langpolicy &-  & vancouver2010.com/code \\
3&vancouver2010.com/forgotpassword & - & canadacode.vancouver2010.com/ user/register \\
4&vancouver2010.com/store & - & canadacode.vancouver2010.com \\
5&vancouver2010.com/store/index.html & - & canadacode.vancouver2010.com/explore \\
6&vancouver2010.com/ & - &  canadacode.vancouver2010.com/ user/login?destination=node/add/image \\
7& canadacode.vancouver2010.com & - & canadacode.vancouver2010.com/pulse \\
8&canadacode.vancouver2010.com/nfb-onf & - & canadacode.vancouver2010.com/challenge \\
9&canadacode.vancouver2010.com/contact & - & i-credible.nl \\
10&canadacode.vancouver2010.com/resources & - &  vpzschaatsteam.nl \\ \hline
\multicolumn{4}{}{} \\ \hline\hline
 
& \textbf{Feb-2010} &\textbf{ Mar-2010} & \textbf{Collection} (\textit{Nov-09 to Mar-10}) \\ \hline\hline
1&monlibe.liberation.fr & monlibe.liberation.fr & monlibe.liberation.fr \\
2&topsport.com/sportch/liveticker/ & laprovence.com/la-provence-le-faq-de-la-moderation & vancouver2010.com/code \\
3& lefigaro.fr &  get.adobe.com/flashplayer &  lefigaro.fr \\
4&laprovence.com/la-provence-le-faq-de-la-moderation & vancouver2010.teamgb.com /teamgb/team-behind-team-gb/filenotfound.aspx & laprovence.com/la-provence-le-faq-de-la-moderation \\
5&lefigaro.fr/sport & ledauphine.com & lefigaro.fr/sport \\
6&get.adobe.com/flashplayer &  lefigaro.fr/economie & get.adobe.com/flashplayer \\
7& lefigaro.fr/meteo &  lefigaro.fr/sport &  lefigaro.fr/meteo \\
8& lefigaro.fr/le-talk &  lefigaro.fr/actualites-a-la-une &  lefigaro.fr/le-talk \\
9& dosb.de/de/vancouver-2010/vancouver-ticker/detail/printer.html &  lemonde.fr/cgv & topsport.com/sportch/liveticker/ \\
10&ledauphine.com & ffs.fr/index.php &  vancouver2010.com/en/langpolicy \\ \hline
\end{tabular}
\end{table*}

\begin{table}[tb]
\caption{Overlapping and (Correlation) between the top 50 results.}
\label{tab:cor}

\begin{tabular}{c|c|c|c|c } \hline
 &  \textbf{Jan} & \textbf{Feb} & \textbf{Mar} & \textbf{Collection}\\ \hline
Nov & 29 (-0.042)& & &27 (0.078)  \\ \hline
Jan  & & 5 (0.089) & &27 (0.169)\\ \hline
Feb  & & &16 (0.053) & 22 (0.185) \\ \hline
Mar  & & & & 14 (0.062)\\ \hline
\end{tabular}

\end{table}

\subsection{Time-Indexed Inlinks}
The Memento protocol provides a TimeMap  (a list of all the available mementos) for each URI-R. For each memento, the user could extract the list of outlinks but this is an expensive operation as seen by Weber \cite{Weber2012} while Thelwall and Vaughan \cite{Thelwall2004} used AltaVista to retrieve the inlink information from the current web instead of IA in order to examine the country balance in IA. With the current WayBack Machine, it is not possible to extract the incoming links for specific URI-R through the time.

Listing \ref{lst:accessschema} shows a sample from the access response for \url{vancouver2010.com}. The response has a complete list of the available incoming links to this URI. From this response, we can produce various information. For example, table \ref{tab:inlinksAnchorText} shows a sample list of the anchor text for different incoming links to \url{vancouver2010.com} ordered by time. 
This is not feasible with the regular WayBack Machine. If the $URI-R$ is not in the archive, this summary of anchor text could provide a good description of what is missing.

\begin{table}[tb]
	\caption{vancouver2010.com Inlinks anchor text.}
	\centering
\begin{tabular}{l p{2.5in}} \hline
\textbf{Date}  & \textbf{Text} \\ \hline
04-Nov-09&vancouver2010.com\\ 
11-Nov-09&vancouver2010.com\\ 
18-Nov-09&vancouver2010.com\\ 
16-Jan-10&Vancouver 2010 Olympic Games\\ 
16-Jan-10&Vancouver 2010 Olympic Games\\ 
23-Jan-10&vancouver2010.com\\ 
23-Jan-10&2010 Vancouver Olympic Games Medals Results Schedule Sports\\ 
30-Jan-10&2010 Vancouver Olympic Games Medals Results Schedule Sports\\ 
30-Jan-10&vancouver2010.com\\ 
30-Jan-10&Vancouver 2010 Olympic Games\\ 
13-Feb-10&Vancouver 2010 Olympic Winter Games\\ 
15-Feb-10&Vancouver 2010 Olympic Games\\ 
18-Feb-10&Official Vancouver Games site\\ 
19-Feb-10&vancouver2010.com\\ 
20-Feb-10&Official Vancouver Games site\\ 
21-Feb-10&VANOC 2010\\  \hline
\label{tab:inlinksAnchorText}
\end{tabular}
\end{table}

\section{Conclusions}

We presented ArcLink, a distributed system to construct, preserve and deliver the temporal web graph for large-scale web archives. ArcLink extends the current Wayback Machine with APIs interface to access the link structure metadata on fine-grained level. ArcLink optimization techniques reduced the input corpus to 29\%, extracted efficiently from the WARC files with 61\% of the regular page scraping, and deliver the link structure in RDF/XML format. ArcLink provided an adequate platform for new applications. Temporal PageRank had weak relation between the rank at each month, and Time-Indexed Inlinks gave information about URI through the time.



\section{Acknowledgments}
This work is supported in part by the Library of Congress and NSF IIS-1009392. We would like to thank Kris Carpenter Negulescu, Aaron Binns, and Vinay Goel from Internet Archive for allowing us to use the IA infrastructure.

\begin{lstlisting}[float=*,caption={Link Structure partial response session.},label={lst:accessschema},numbers=left,stepnumber=5,numberstyle=\tiny ,language=xml,morekeywords={rdf:Description,twg:hasOutlinks,twg:hasInlinks}]
> curl "http://localhost:8080/LinkService/linkQuery?uri=vancouver2010.com"

<?xml version="1.0"?><rdf:RDF xmlns:rdf="http://www.w3.org/1999/02/22-rdf-syntax-ns#"
xmlns:twg="http://www.mementoweb.org/TemporalWebGraph/"> 
<rdf:Description rdf:about="vancouver2010.com">
 <twg:hasOutlinks rdf:parseType="Collection">
   <rdf:Description rdf:about="paralympic-games)/news">
    <twg:type>href</twg:type> <twg:text>News</twg:text>
    <twg:timestamp>  <rdf:Bag>
     <rdf:li>20091103011307</rdf:li><rdf:li>20100130003005</rdf:li> ...
    </rdf:Bag> </twg:timestamp> </rdf:Description>
   <rdf:Description rdf:about="olympic-cross-country-skiing)/">
    <twg:type>href</twg:type> <twg:text>Cross-Country Skiing</twg:text>
    <twg:timestamp> <rdf:Bag>
     <rdf:li>20091110011557</rdf:li> <rdf:li>20100227081100</rdf:li>  ...
    </rdf:Bag> </twg:timestamp> </rdf:Description>
.....
 </twg:hasOutlinks>
 <twg:hasInlinks rdf:parseType="Collection">
   <rdf:Description rdf:about="http://vancouver2010.teamgb.com/gallery/gillian-cooke/">
    <twg:type>href</twg:type> <twg:text>Official Vancouver Games site</twg:text>
    <twg:timestamp> <rdf:Bag>
     <rdf:li>20100217101229</rdf:li>
    </rdf:Bag> </twg:timestamp> </rdf:Description>
     <rdf:Description rdf:about="http://www.swissolympic.ch/olympiablog/?tag=/verletzung">
     <twg:type>href</twg:type> <twg:text>VANOC 2010</twg:text>
    <twg:timestamp> <rdf:Bag>
     <rdf:li>20100220104902</rdf:li>
    </rdf:Bag> </twg:timestamp> </rdf:Description>
....
 </twg:hasInlinks>
</rdf:Description></rdf:RDF>
\end{lstlisting}

\begin{small}

\bibliographystyle{abbrv}
\bibliography{../../Bibtex/library}
\end{small}

\balancecolumns
%
%
%


\end{document}